\documentclass{article}
\usepackage[utf8]{inputenc}
\usepackage{authblk}
\usepackage{graphicx}
\usepackage{color}
\usepackage{geometry}
\usepackage{hyperref}
\usepackage{lipsum}
\begin{document}

\title{Effectiveness of dismantling strategies on moderated vs. unmoderated online social platforms}

\author{Oriol Artime$^1$, Valeria D'Andrea$^1$, Riccardo Gallotti$^1$, Pier Luigi Sacco$^{2,3,4}$, Manlio De Domenico$^1$}
\affil{$^1$CoMuNe Lab, Fondazione Bruno Kessler, Via Sommarive 18, 38123 Povo (TN), Italy.}
\affil{$^2$Fondazione Bruno Kessler, Via Santa Croce, 77, 38122 Trento, Italy.}
\affil{$^3$IULM University, Via Carlo Bo, 1, 20143 Milan, Italy.}
\affil{$^4$Berkman-Klein Center for Internet \& Society, Harvard University, 23 Everett St \# 2, Cambridge MA 02138 USA.}

\date{\today}

\maketitle

\begin{abstract} 
Online social networks are the perfect test bed to better understand large-scale human behavior in interacting contexts. Although they are broadly used and studied, little is known about how their terms of service and posting rules affect the way users interact and information spreads. Acknowledging the relation between network connectivity and functionality, we compare the robustness of two different online social platforms, Twitter and Gab, with respect to dismantling strategies based on the recursive censor of users characterized by social prominence (degree) or intensity of inflammatory content (sentiment). We find that the moderated (Twitter) vs un-moderated (Gab) character of the network is not a discriminating factor for intervention effectiveness. We find, however, that more complex strategies based upon the combination of topological and content features may be effective for network dismantling. Our results provide useful indications to design better strategies for countervailing the production and dissemination of anti-social content in online social platforms.
\end{abstract}

\section{Introduction}
Online social networks provide a rich laboratory for the analysis of large-scale social interaction and of their social effects~\cite{szell2010multirelational, kramer2014experimental,borge2016dynamics,lorenz2019accelerating}. They facilitate the inclusive engagement of new actors by removing most barriers to participate in content-sharing platforms characteristic of the pre-digital era~\cite{sacco2018culture}. For this reason, they can be regarded as a social arena for public debate and opinion formation, with potentially positive effects on individual and collective empowerment~\cite{wellman2001does}. However, some of the structural and functional features of these networks make them extremely sensitive to manipulation~\cite{bradshaw2017troops,stella2018bots,stella2019influence}, and therefore to anti-social or plainly dysfunctional influencing aimed at fueling hate toward social groups and minorities~\cite{awan2014islamophobia}, to incite violence and discrimination~\cite{ben2016hate}, and even to promote criminal conduct and behaviors~\cite{muller2019fanning}. As such platforms have been functioning for a very short time on a historical scale, our experience with them is still limited, and consequently our awareness of their critical aspects is fragmented~\cite{hjorth2019understanding}. As a consequence, the governance of such platforms is carried out on a trial-and-error basis, with problematic implications at many levels, from the individual to the institutional~\cite{linke2013social, denardis2015internet}. The most fundamental issue is the lack of widely agreed governance principles that may reliably address the most urgent social challenges~\cite{kaun2011divergent}, and this is a consequence of the fact that we are still at the beginning of the learning curve. On the one hand, we are becoming increasingly aware that certain features of human social cognition that were culturally selected in a pre-digital era may call for substantial and quick adaptations to tackle the challenges of massive online social interaction~\cite{dunbar2012social, meshi2015emerging}. On the other hand, the massive potential public exposure that characterizes such interactions may reinforce typical human biases that lead to people's self-embedding in socially supportive environments~\cite{lee2017homophily}, with the formation of echo chambers built on the assortative matching to, and reciprocal support from, like-minded people~\cite{geschke2019triple, baumann2020modeling}. Moreover, the presence of strong incentives to the acquisition of visibility and social credibility via the building of large pools of followers, and the strong competition for limited attention tend to favor the production and selective circulation of content designed to elicit strong emotional arousal in audiences~\cite{brady2017emotion}, rather than to correctly report information and reliably represent facts and situations~\cite{vargo2018agenda}, with the well-known proliferation of fake news and more generally of intentionally manipulative content~\cite{alrubaian2017reputation, lazer2018science}. A further critical element is the possibility to disseminate across such platforms purposefully designed artificial agents whose goal is that of generating and amplifying inflammatory content that serves broader strategies of audience manipulation~\cite{howard2017junk}, conceived and undertaken at higher levels of social agency by large sophisticated players that may reflect the agendas of interest groups, organizations, and governments~\cite{johnson2019hidden}.

On the basis of such premises, it becomes very important to understand the relationship between the structural features of such networks and their social effects, for instance in terms of manipulation vs. effective control of the production and dissemination of anti-social and inflammatory content. In this perspective, two main approaches have developed: one that relies upon extensive regulation of user-generated content~\cite{langvardt2017regulating}, and one that basically leaves this task to the community itself, entrusting the moderation of content production and dissemination to the users ~\cite{lampe2014crowdsourcing}. A meaningful question that arises in this regard is: which approach is more desirable as it comes to countering anti-social group dynamics? If socially dysfunctional content is being generated and spread across a certain online social platform, in which case can this be handled more effectively through a targeted removal from the network of problematic users? And how does this change when the network is moderated, and when it is not? Quite surprisingly, at the moment the existing literature does not provide us with an answer to these fundamental question,  and this powerfully illustrates how fragmented and precarious our understanding of such phenomena still is. 

Online social networks often represent typical examples of complex social systems~\cite{onnela2010spontaneous}, and therefore lend themselves to be studied and analyzed through the conceptual toolbox of network science~\cite{hu2008analysis}. This is the approach that we follow here, where we comparatively analyze two different online social platforms, Twitter and Gab, from the point of view of the robustness with respect to different types of `attacks', namely of targeted removals of some of their users from the respective networks. This kind of analysis may be seen as an experiment in understanding the effectiveness of alternative strategies to counter the spreading of socially dysfunctional content in online social platforms, thereby providing us with some first insights that may be of special importance in designing the governance of current and future platforms. The reason why we chose Twitter and Gab for our study is that, primarily, they clearly reflect the two main options we want to test in our analysis: Twitter is a systematically moderated platform, whereas Gab is essentially unmoderated~\cite{zannettou2018gab, kalmar2018twitter}. On the other hand, the two platforms are very similar in many other features, and this makes of them a good basis for a comparison, as opposed to other online social platforms with substantially different features. In addition to this, data on these platforms can be collected with relative ease, and this allows the construction of scientifically sound databases more easily than it is the case with other platforms.

The implementation of the degree-based dismantling strategy, which turns out to be a very effective one, shows that, surprisingly, the moderated or unmoderated nature of the networks does not induce clear robustness patterns. To understand the way the different networks respond to this type of intervention we resort to higher-order topological correlations: the degree assortativity and a novel concept, the inter-$k$-shell correlation, which is designed to give an idea of how the different centrality hierarchies in the network are connected among them. We complement our analysis by proposing dismantling strategies based on the sentiment of the users, aiming at evaluating how robust are the networks under the removal of a certain type of users. We find that the networks are indeed quite robust, hence global communicability is guaranteed, if one blindly removes the users based only on their sentiment. For a faster network dismantling, we propose strategies that combine topological and sentimental information.

\section{Results}

\subsection{Main features of the datasets}

To analyze the robustness of Twitter vs. Gab to targeted attacks, that is, the structural consequences of the removal of certain users from the network according to specific criteria, we had to select a topic that would provide the criterion to identify users in terms of their participation in the online debate, of the inflammatory content of their posts, of their structural position in the online community, and so on. We chose to focus on posts referring to the current president of the United States of the America, Donald Trump, and containing the following keywords: `trump', `potus'. The choice of Donald Trump as the reference topic is due to the high salience and popularity of this public figure in online social media debates, which are typically characterized by highly polarizing and inflammatory content, thus making of it an ideal test bed for our analysis~\cite{eddington2018communicative}.

Content on Gab on a time window of 3 months in 2018, gathering a total of almost 450k posts, has been made publicly available from \url{https://pushshift.io/}. As Twitter is much more populated than Gab and characterized by much higher volumes of online activity, here we collected content in a time window of two months in 2018, gathering a total of almost 45M posts. More detailed information and technical specifications may be found in the \hyperref[sec:Methods]{Methods} section.

To reconstruct the behavioral networks from the online social platforms, we represent users as nodes and interactions between users as links of a social network. For both Twitter and Gab, we considered two different types of networks, capturing different facets of social interactions. The first one is replies: whenever a user replies to the message of another user, a link between the two respective nodes is established. The second one is mentions: whenever a user mentions another user in one message, a link between the two respective nodes is established. For the sake of simplicity, we consider the networks as un-directed --- no information is encoded about who is the sender and who is the receiver of the messages ---, unweighted --- no information about the number of interactions ---, and time-aggregated --- no information about the timing of the interactions. 

Replies and mentions correspond to two different aspects of social interaction~\cite{greenhow2012twitteracy}. In the case of replies, users are engaged in an active conversation between them, which can also be unilateral if one user responds but the other does not in turn. In the case of mentions, one user is pointing attention of other users toward a third user, but the mentioning and mentioned users need not be engaged in a direct conversation between themselves. Replies are therefore part of a dyadic interaction as in a typical conversation (where clearly one user may establish several conversations at the same time if multiple other users reply to his/her posts), whereas mentions are typically part of a multi-lateral conversation that may intrinsically involve several users at the same time, and even be targeted to reach an indefinite number of users.

On this basis, it is possible to construct analogous networks for Twitter and Gab users replying to, or mentioning, other users in messages that contain Donald Trump related hashtags. The purpose of our analysis is to investigate to what extent such networks are resilient to different types of attacks consisting of the removal of a number of users with specific characteristics, on the basis of the different nature and characteristics of the two online social networks in terms of moderation.

\subsection{Robustness and topological properties}

\begin{figure}
    \centering
    \includegraphics[width=1.\textwidth]{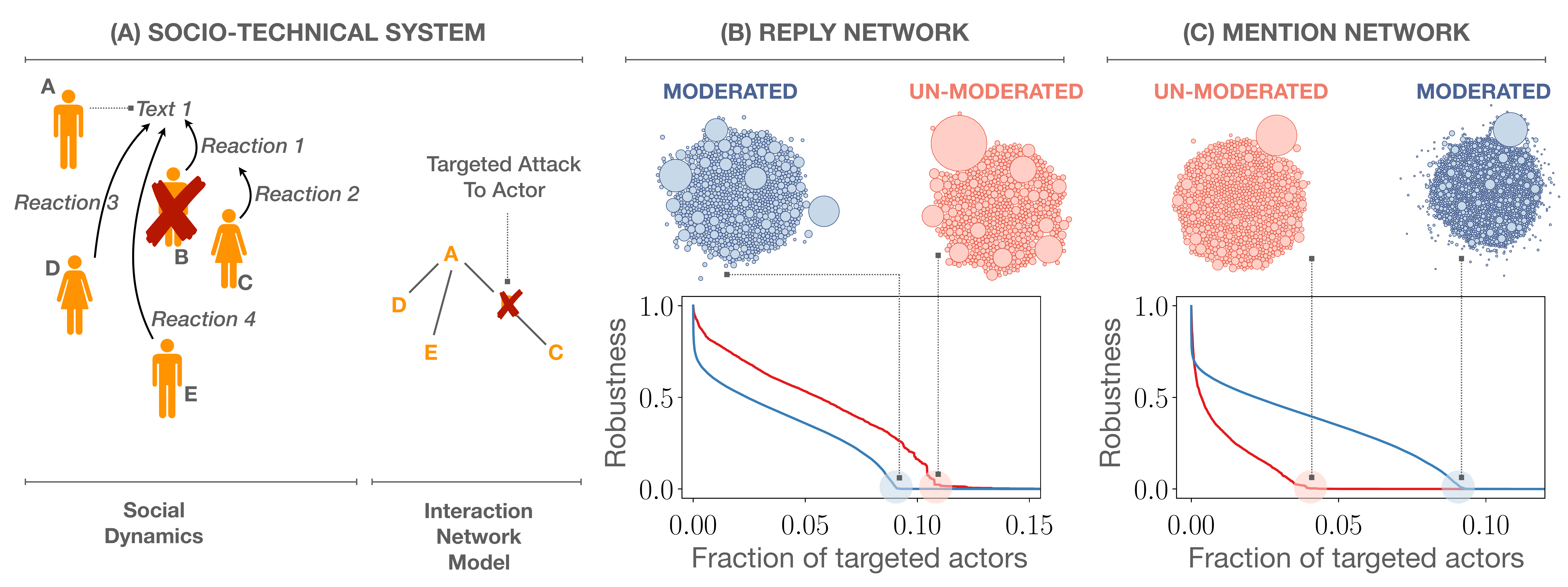}
    \caption{(\textbf{A}): Sketch showing the mapping between the online dynamics and the reconstructed behavioral networks. Attacks are identified with the removal of agents and might split the network in smaller components. In our case, two sub-networks appear once the actor $B$ is attacked: $\{ A, D, E\}$ and $\{ C \}$. The overall robustness of a system is related to its capacity of maintaining a component as large as possible when suffering attacks. (\textbf{B}) and (\textbf{C}): Curves of the largest connected component under degree-based attacks in the network of replies and the network of mentions, respectively. ``Moderated'' corresponds to Twitter and ``unmoderated'' to Gab. The above bubble plots show the connected components remaining at the percolation point. The size of the bubbles is related to the logarithm of the component size.
    }
    \label{fig:fig1}
\end{figure}

We study the robustness of the reconstructed networks by means of percolation theory~\cite{stauffer1991introduction}. The basic procedure is to delete nodes according to a given criterion and, as the process unfolds, compute several properties of the damaged system. In the context of online social networks, node deletion can be identified, as already hinted, with the banning or temporary inhibition of a specific user, see Figure~\ref{fig:fig1}. A good topological proxy to assess the robustness of the network is the largest connected component (LCC), which is the largest connected sub-network remaining after user removal. When the normalized size of the LCC, $S$, is close to $0$, the network is completely disintegrated in many small clusters, and therefore there is no possibility to observe propagation of information at a global scale. On the contrary, when $S$ is close to $1$, the removal of nodes barely affects the overall topology, and information can potentially flow between almost every pair of actors. The passage from $S \neq 0$ to $S = 0$ is called the percolation phase transition~\cite{dorogovtsev2008critical}, and the exact value of the fraction of removed nodes for which the size of the LCC becomes null is called the percolation point~\cite{PercolationClarification}. Roughly speaking, a network is considered robust when it can handle a large amount of node removals without being disintegrated. 

The most basic procedure to study network robustness is the random removal of nodes, which is equivalent to the classic percolation process with degree heterogeneity~\cite{callaway2000network}. Random attacks are known to be a poor strategy to break a network, and especially when the degree distribution is broad~\cite{albert2000error, cohen2000resilience}, which is a hallmark of social networks~\cite{newman2018networks}. This is because random node selection will pick with high probability low-degree nodes, which do not play any significant role in keeping the network connected. A more effective strategy to destroy a network consists in selecting nodes by degree, and remove those with the highest degree first. In this case, the percolation point is significantly reduced~\cite{albert2000error, cohen2001breakdown}. This response ---weak to targeted attack but strong to random failures--- has been dubbed the robust-yet-fragile effect~\cite{doyle2005robust}. 

As an alternative to random attacks, here we perform degree-based attacks in our reconstructed networks (Figures~\ref{fig:fig1}B and \ref{fig:fig1}C), thus targeting structurally prominent users rather than randomly picked ones. To implement such attacks, we use an adaptive scheme: after each removal, we recalculate the degrees and recursively choose the node with the highest degree to be deleted next. For all networks we find, as expected, a relatively low percolation point, i.e., low robustness, a result that agrees with the idea that actors in social networks are heterogeneously connected among each other. Another quantity that is heterogeneously distributed is the size of the connected components at the percolation point. Their sizes are shown as bubble plots in Figures~\ref{fig:fig1}B and \ref{fig:fig1}C, where the radius of each bubble is proportional to the logarithm of the number of actors in the components. The cluster size probability density function is shown in Figure~\ref{fig:figSM1}. These plots indicate that at the precise moment the communication in our social networks cannot be held global anymore, it is very likely to find large clusters with sizes that are very far away from the mean cluster size. Indeed, the distribution turns out to be power-law $ p(s)\propto s^{-\alpha}$, with exponent $\alpha$ smaller than $3$, a signature of infinite variance. This property only holds at the percolation point, whereas away from it the finiteness of the variance is recovered due to an exponential cutoff in the component size distribution. 

\begin{figure}
    \centering
    \includegraphics[width=1.\textwidth]{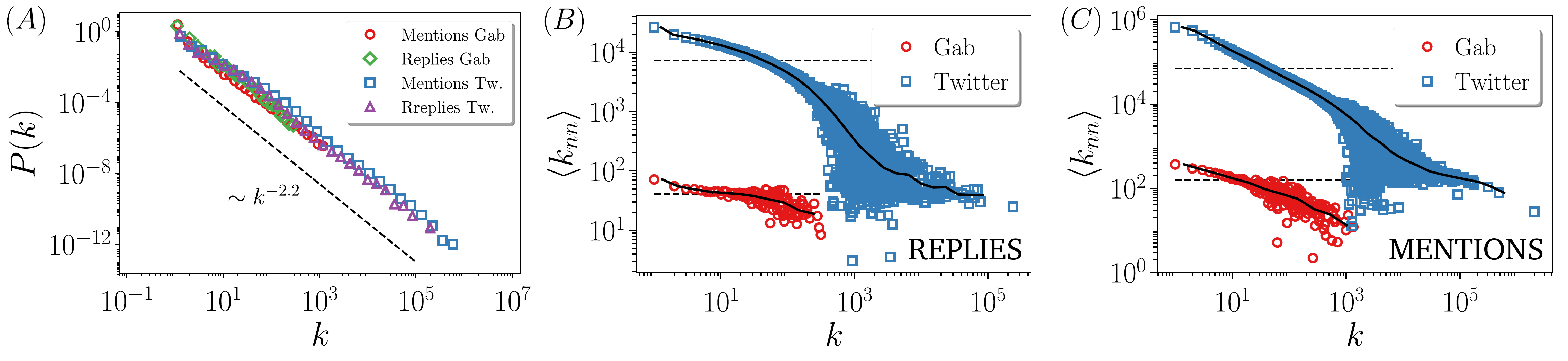}

    \caption{(\textbf{A}): Degree distribution for the different networks and a curve scaling as $k^{-2.2}$ to guide the eye. Degree correlations for the networks of replies (\textbf{B}) and for the networks of mentions (\textbf{C}). The solid curve is computed by averaging the data within $k$-bins, and the dashed horizontal line corresponds to $\langle k^2 \rangle / \langle k \rangle$, which is the value that $\langle k_{nn} \rangle$ would take if correlations are washed out.}
    \label{fig:fig2}
\end{figure}

We observe that the presence or absence of content control in the online social network does not induce a clear robustness pattern. Indeed, in the network of replies the percolation point for Gab is larger than the one for Twitter, whereas in the network of mentions the result is reversed. This leads us to conclude that content policies cannot be directly correlated to robustness assessments in online social networks. To better understand the response to degree-based attacks, we need to shed light on the topological properties of the networks. The first thing to note is that since for uncorrelated networks, i.e., those networks where the degree of an actor is independent of her neighbor's degree, the percolation point is known to only depend on the first and second moment of the degree distribution~\cite{molloy1995critical}, and the degree distribution for all networks we considered is very similar (Figure~\ref{fig:fig2}A), the reason for the variability in the robustness must therefore be hidden in topological correlations. A topological correlation that is known to affect the percolation point is the mixing assortativity~\cite{newman2002assortative}, known as homophily in the social sciences~\cite{mcpherson2001birds}, which reflects the tendency of actors to interact with peers with similar characteristics. At the topological level we have the degree assortativity, which is positive when nodes are mainly connected to other nodes of similar degree, and is negative when the opposite is found. A broadly used measure to capture the level of homophily is the assortativity coefficient~\cite{newman2018networks}
\begin{equation}
    r = \frac{\langle q q' \rangle_l - \langle q \rangle_l \langle q' \rangle_l}{\langle q^2 \rangle_l - \langle q \rangle^2_l},
    \label{eq:e1}
\end{equation}
where $\langle \cdot \rangle_l$ denotes average over all links, and $q$ and $q'$ are the excess degrees of the nodes at the end of a link, that is, the number of edges attached to a node other than the one we reached out to. $r$ is nothing else than the Pearson correlation coefficient of the degrees at either end of an edge, and is normalized within the interval $[-1, 1]$. The assortativity coefficient is appealing because it encodes the correlations into a single number, which is usually employed to infer the robustness of the network. The percolation point of a network with positive (negative) assortativity is higher (lower) than the percolation point of a network with $r=0$, assuming the same underlying degree distribution in all cases~\cite{newman2003mixing}. The larger the assortativity coefficient (in absolute value), the greater the separation with respect to the percolation point of $r = 0$. In our case, we find that for the network of replies $r_{gab} = -0.16 < r_{twitter} = -0.05 $ and for the network of mentions $r_{gab} = -0.26  < r_{twitter} = -0.04 $~\cite{AssortativityClarification}. The values of the assortativity coefficient agree well with the robustness patterns in the networks of mentions, but not with those in the network of replies. Hence, the information brought by $r$ is not sufficient to successfully explain the response of our system to degree-based attacks.

The implicit assumption behind the assortativity coefficient is that the degree correlation ---the mean degree of the neighbors of all degree-$k$ nodes $\langle k_{nn}(k) \rangle = \sum_{k'} k' P(k'|k)$, where $P(k'|k)$ is the conditional probability that following a link of a $k$-degree node we arrive at a degree-$k'$ node---  has a linear dependence on the degree, with slope $r$. If $\langle k_{nn}(k) \rangle \sim r k $ does not apply, the value (and even the sign) of $r$ can be misleading, as well as the correct interpretation of the location of the percolation point as a function of $r$. We show in Figures~\ref{fig:fig2}B and \ref{fig:fig2}C that the linear assumption does not hold, although in all cases we observe a monotonically decreasing tendency, concluding that all networks are dis-assortative. One could argue that in the network of replies, Twitter's $\langle k_{nn}(k) \rangle$ decays faster than in Gab, thus ensuring a higher robustness to the latter. This very same argument, though, fails when applied to the network of mentions. 

\begin{figure}
    \centering
    \includegraphics[width=1.\textwidth]{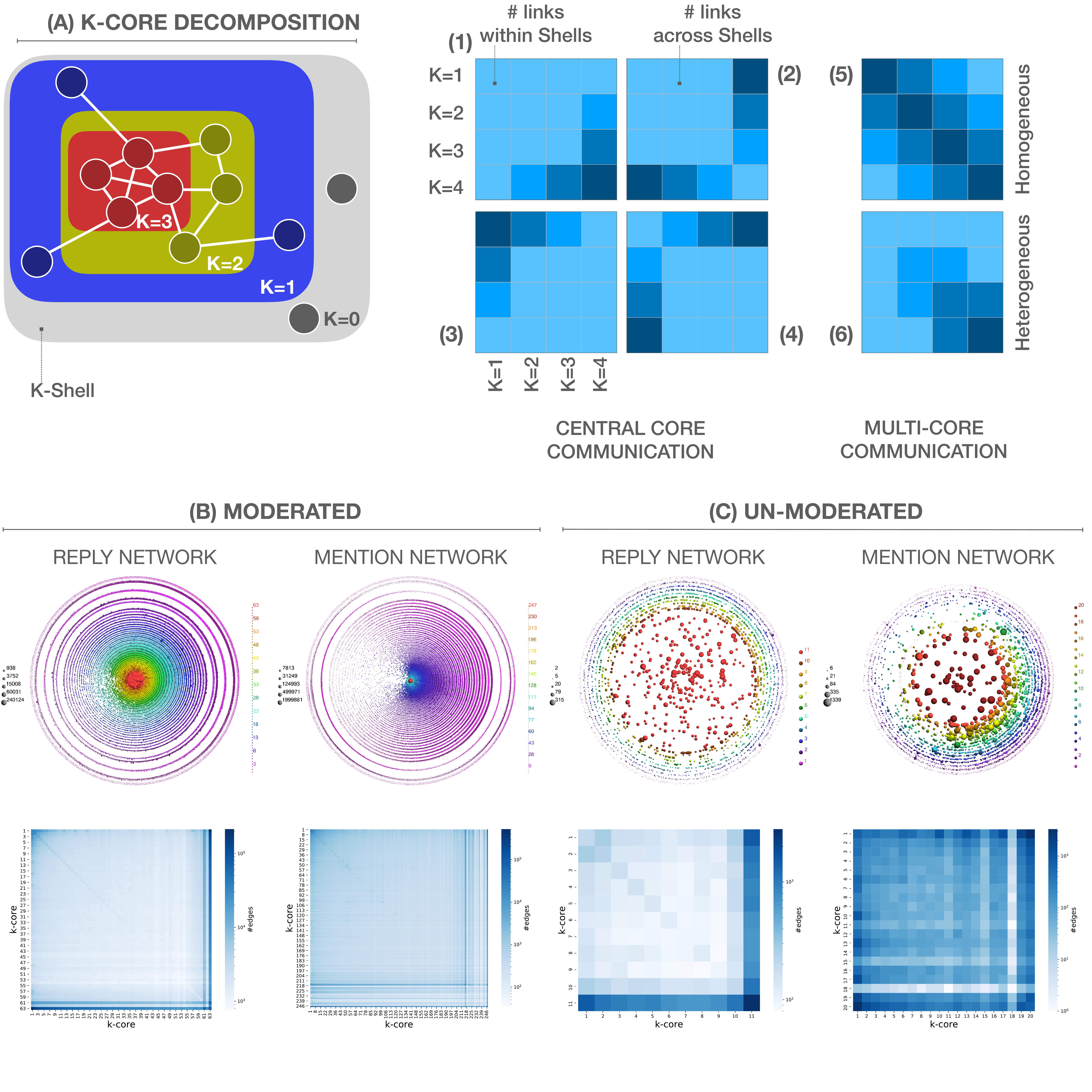}
    \caption{(\textbf{A}): Sketches showing the $k$-core structure of a toy network and heat maps with a decentralized and centralized network organization. 
    The patterns exemplified in the heat maps A1 to A4 represent scenarios in which the connectivity is involving predominantly a single $k$-shell. In A1 most connections are between nodes of the central core ($K=4$) and other central shells (as in assortative networks). In A2 most connections are between the central core and nodes of external shells (as in dis-assortative networks). In A3 and A4 most connections involve the most external shell ($K=1$). This can also happen in an assortative way (A3), where most communication will be between isolated couples of nodes of the external shells, or a dis-assortative way (A4) where most communication is between a large external shell of leaves and the central shells (core-periphery structure). 
    The patterns of heat maps A5 and A6 describe instead networks where connections are distributed between multiple shells in such a way that nodes in a $k$-shell are mostly connected to the same shell or to shells below and above in the hierarchy. They can be obtained by overlapping single shell effects similar to that of A1. In A5, connections are homogeneously distributed across shells, while in A6 connections are more abundant in the central shells.
    In (\textbf{B}) and (\textbf{C}), $k$-shell decomposition and associated heat map for Twitter and Gab, respectively. The distribution of nodes and their degrees across the different $k$-shells  can be appreciated by representing the different nodes in concentric circles (plotted above using LaNet-vi~\cite{alvarez2005large}).
    In the heat maps we focus instead on the connectivity. In the light of the pattern exemplified above, we observe how the Twitter reply and mention and the Gab reply networks can be seen as an overlap of A1 and A2 (with the A3 patterns also playing a minor role in Twitter). The Gab mention network differs notably, and can be seen as an overlap of the A2, A4 and A6 patterns: more communication is thus seen in the intermediate shells (A6) and between marginal and central nodes (A4).
    } 
    \label{fig:fig3}
\end{figure}

At this point we must consider higher-order topological correlations to have a satisfactory explanation of why the mention networks do not respond as expected from the assortative profiles. To this purpose, we use the $k$-shell decomposition~\cite{alvarez2005large}, which conveys information about the hierarchical structure of networks. The concept of $k$-shell is intimately related to the one of $k$-core~\cite{seidman1983internal, seidman1983network}, which is the sub-network that survives after removing all nodes with degree less than or equal to $k$ from the original network, see Figure~\ref{fig:fig3}A. The $k$-shell is the subset of nodes that belong to the $k$-core but not to the ($k+1)$-core. A network representation in $k$-shells offers a qualitative way to assess the connectivity and clustering inside $k$-shells, and the degree-shell correlation, i.e., how hubs are central with respect to their $k$-coreness, see~\cite{alvarez2006lanet} for further details. The most important information for our discussion is the fact that hubs, indicated by larger markers in the first rows of Figures~\ref{fig:fig3}B and \ref{fig:fig3}C, are mainly located in the largest $k$-shell for both Twitter networks, and in the replies of Gab. For Gab mentions, on the contrary, there is a larger density of hubs across $k$-shells (first row of Figure~\ref{fig:fig3}B). The presence of hubs in different shells denotes a low degree-coreness correlation; put otherwise, there is an important number of highly connected nodes at the periphery of the network. 

In order to quantitatively support the qualitative observation of the degree-coreness correlation, we propose here to inspect the connections among $k$-shells. Let $K$ be the maximum $k$-shell index. We can compute a symmetric matrix of dimension $K \times K$, whose elements $(i,j)$ are the number of links between nodes of $k$-shells indexes $i$ and $j$. By plotting this matrix as a heat map different patterns can emerge, and they can be interpreted as an indicator of the tendency of nodes to communicate with their peers of same level of $k$-coreness, hence giving a global idea of how centralized is the communication in the network that cannot be captured by the assortativity or degree correlations. See Figure~\ref{fig:fig3}A for further details, where we distinguish among possible behaviors. Thus, we call centralized networks those networks in which most of the nodes in low and intermediate $k$-shells are connected only to those nodes in the largest $k$-shell. Likewise, a decentralized network is characterized by nodes of all $k$-shells connected among them with a similar density of links. We plot in the last row of Figures~\ref{fig:fig3}B and ~\ref{fig:fig3}C such heat maps. In both Twitter networks and in Gab replies we find that almost all inter-$k$-shell interactions occur between the nodes in the largest $k$-shell and the others, i.e., they are centralized. Surprisingly, the heat map of Gab mentions is much more homogeneously populated, indicating that each $k$-shell has a significant portion of links to distribute, and that this distribution covers all the other shells with no particular preference, i.e., the network is decentralized. In light of this inter-$k$-shell connectivity, we can explain why the Twitter mention network is more robust than the Gab one. Since in the former all large-degree nodes are very interconnected among them in the largest $k$-core, deleting them leads to a slow disintegration of the network. In the latter case, the hubs are not so well compacted in the largest $k$-core, but decentralized across different levels in the topological hierarchy, therefore deleting them by degree leads to an easier network disintegration.

\subsection{Sentiment-based attacks}

So far we have analyzed the robustness of the Twitter vs. Gab networks to attacks that are based on the degrees of users in the online network, that is, on targeting the most relationally prominent individuals. Attacks of this kind test the topological properties of the network, and they are known to effectively dismantle the underlying graph, but are at the same time totally blind to any users metadata. However, there can be situations, and this is especially the case in the context of social networks, in which we would like to assess the response of the system to the removal of certain type of users, e.g., fake news generators or hate spreaders. In this section we discuss the effects of carrying out attacks based on the sentiment of the users, as obtained from the body of the messages they write. We will consider only the network of replies in this case, because it has a much higher number of active users with well-defined sentiment than the network of mentions. We characterized the amount of user's text emotional content by a text-based analysis that distinguish three different components of emotions, namely valence, arousal and dominance. We leave for the Supplementary Material the same analysis for other three sentiment classifiers, 
where we reach identical conclusions to the ones presented below.

\begin{figure}
    \centering
    \includegraphics[width=1.\textwidth]{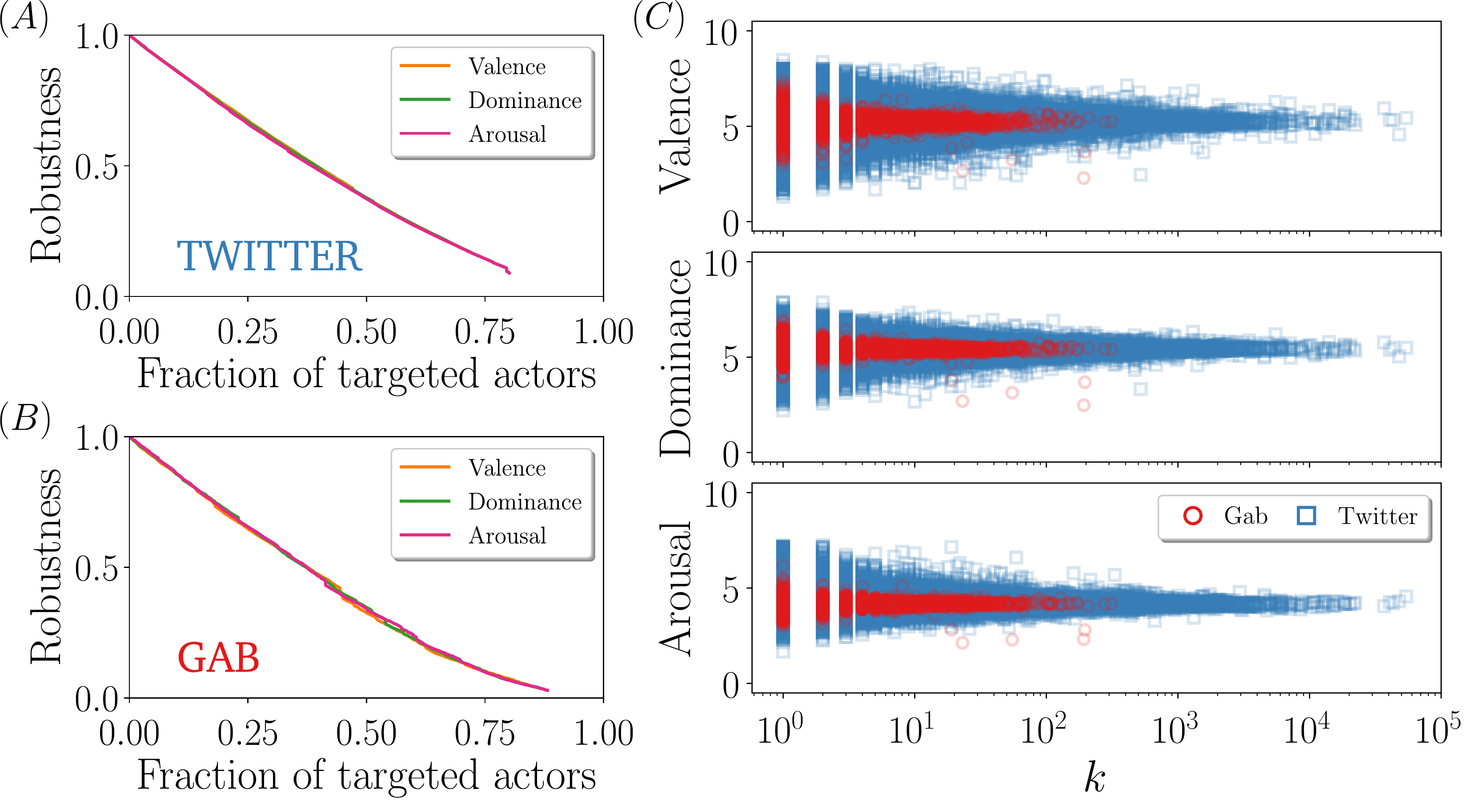}

    \caption{Size of the largest connected component for the sentiment-based attacks in Twitter (\textbf{A}) and in Gab (\textbf{B}). Each curve corresponds to the sentiment targeted for the removal. (\textbf{C}): Correlation between the sentiment score and the degree of the nodes. The corresponding sentiment is written in the vertical axes. For the sake of clarity, in Twitter networks we only plot a random $20\%$ selection of all users.}
    \label{fig:fig4}
\end{figure}

We perform the attacks by sorting nodes with decreasing intensity of the (numerical indicators of the) different sentiments, and deleting the nodes following that order. In other words, at each round we delete those users whose posts are the most emotionally charged (i.e., potentially most inflammatory) among those still present in the network. We display the results of this process in Figures~\ref{fig:fig4}A and \ref{fig:fig4}B. The most striking result is that the percolation point is quite large, that is, one needs to remove practically all actors to break apart the main component. Moreover, the percolation curves are similar to the typical response of a scale-free network subject to random attacks, therefore indicating that either most of the extreme sentiments are located in the low-degree regime, or that, at least, there is no significant correlation between the position of a user in a network and the sentiment of the messages s/he writes.

Figure~\ref{fig:fig4}C sheds some light on the interplay between topology and sentiment. We can observe that, indeed, the most extreme sentiments values are located in the area of low degrees, which explains why the sentiment-based attacks do not result in a very effective dismantling of the networks. Another particularity of the sentiment-topology correlation is that the sentiment score becomes independent of the degree as $k$ grows. This is somewhat surprising, as one would expect non-trivial effects of the network structure on the sentiment distribution. A plausible explanation of this result is that, as users become more active (larger degree), their sentiment values, which are computed over all their written posts, average out resulting in a neutral sentiment value.

\begin{figure}
    \centering
    \includegraphics[width=0.9\textwidth]{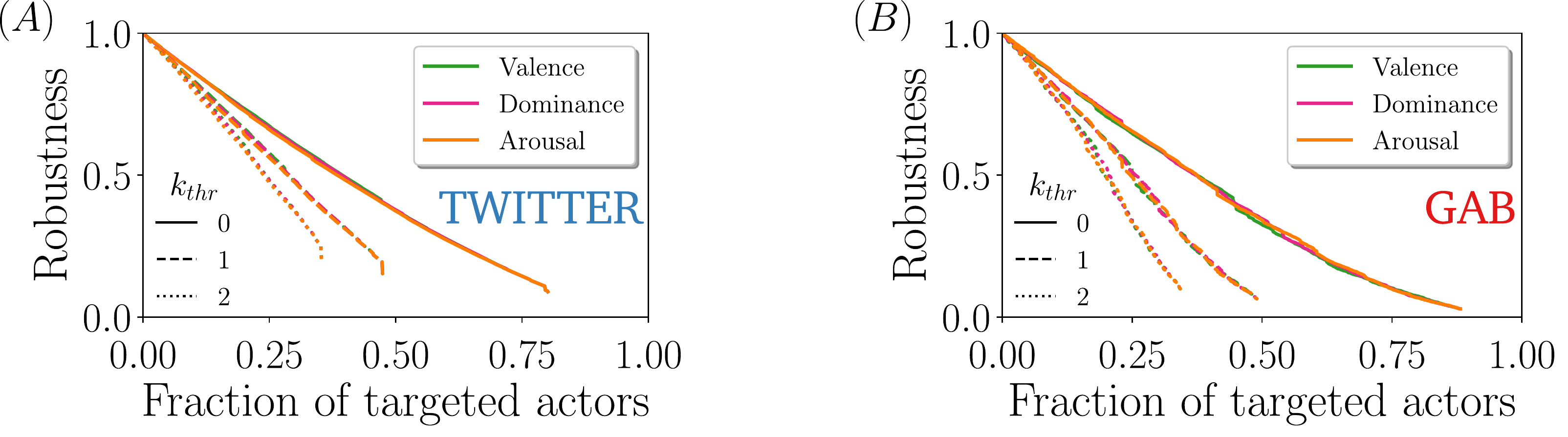}
    
    \caption{Size of the largest connected component for sentiment-based attack applied to nodes with degree $k>k_{thr}$ in Twitter (\textbf{A}) and Gab (\textbf{B}). Line color corresponds to the attacked sentiment and the line style to the low-degree threshold. Note that the $S$ curves do not arrive at $0$ because the gap reflects the nodes that are not slated for deletion (either $k \leq k_{thr}$ or they do not have an assigned sentiment).}
    \label{fig:fig5} 
\end{figure}

An important consequence of this particular sentiment-topology correlation is that when it comes to protect a network against certain type of content or type of users, blind removal proves to be a very efficient strategy at a global level, in the sense that guarantees the potential exchange of information between all the remaining pair of nodes. However, if the goal is to use sentiment information to achieve an efficiently network dismantling, the implemented strategy must be combined with some topological information. One of the options is to sort from highest to lowest the users according to a sentiment, but ignoring those below a certain degree threshold $k_{thr}$. That is, for example, remove users that generate hateful content as far as these users are active above a certain threshold. With this simple modification, we see a significant decrease in the percolation point, i.e., the networks become less robust. The percolation curves for different values of $k_{thr}$ are shown in Figure~\ref{fig:fig5}. We indeed see that passing from $k_{thr} = 0$ to $1$ already reduces the percolation point by about $45\%$ in Gab and $40\%$ in Twitter, while passing from $k_{thr} = 0$ to $2$ leads to a reduction of about $61\%$ in Gab and $56\%$ Twitter.

\section{Discussion}

In scholarly and policy debates, great emphasis is placed on the issue of whether online social platforms should be moderated or not~\cite{gillespie2018custodians}. Clearly, one of the main issues is to what extent moderation can be effective in leading users to a more responsible use of the platform and to keep anti-social attitudes and contents under control~\cite{myers2018censored}. One would expect that the presence of moderation or the lack thereof would make a big difference in terms of the effectiveness of network dismantling interventions against the production and diffusion of dysfunctional content promoting hate, violence, discrimination, and the like. However, a somewhat surprising result of our analysis is that moderation does not seem to be a crucial factor to consider in the assessment of the effectiveness of network dismantling, and therefore in evaluating which type of online social platform may be more easily governed in this regard.

Looking into the results in more detail gives some extra insight. We have considered two different kinds of networks, one based upon replies to posts by another user, and the other upon mentions of another user in the body of text of someone's post. In terms of anti-social content, the replies network is typically bound to capture the dysfunctional evolution of online conversations (e.g., flames), whereas the mentions network is more geared toward capturing invitations to group-based attacks (e.g. shaming). What we find is that, when considering dismantling strategies based upon topological network features (i.e. users' degrees), Gab proves to be more robust in the replies network whereas Twitter is more robust in the mentions network. In unmoderated networks, group-based attacks to users may rapidly escalate and become particularly virulent, and therefore removing those haters who are most connected may effectively power down the circulation of information across the network. In the case of a moderated platform, such escalation is partly already filtered out and so the attack could turn out relatively less effective.  This seems to be a good reason why Gab proves to be less robust than Twitter in this context. On the other hand, when moderation is present in the network, group-based attacks are more easily detected and filtered out, whereas dysfunctional conversations are more elusive and can more easily survive the filter. When the filter is relatively less effective, network breakdown may make the difference. In this case, removing the most connected users may dismantle an information flow that would be difficult to block otherwise, and this is a possible reason why Twitter is relatively more affected by attacks to the replies network.

Considering higher-order topological correlations, and therefore the hierarchical structure of networks by means of $k$-shell decomposition, it turns out that the network hubs tend to be located in hierarchically central parts of the network in the case of both Twitter networks and in the replies network of Gab, In the mentions network of Gab we find instead a different structure where network hubs are more distributed and therefore also marginal areas of the network maintain high levels of connectivity. Even if the global network architecture may be dismantled, therefore, in the Gab mentions network the marginal parts may remain highly active and cohesive nevertheless. In this case, therefore, the dismantling of global information flow cannot be regarded as a fully satisfactory outcome in terms of network attack. In particular, a substantial risk remains that once the core users have been successfully removed, previously marginal pockets that survive the attack may subsequently gain more centrality in the remaining network and launch a new cycle of dysfunctional content creation and propagation. 

One could consider, however, a more intuitive way of designing and carrying out the attacks ---namely, directly targeting the users who create and spread the most inflammatory content. We have therefore analyzed, for the replies network, an alternative strategy of sentiment-based attacks where the emotional charge of posted content may be evaluated with respect to the scale of a specific target sentiment. The users who rank higher in sentiment activation may therefore be recursively targeted and removed. However, this alternative strategy seems to be less effective, so that actual network dismantling calls for a high number of cycles of removal that may practically amount to break down the whole network. The reason for this is intuitive: users who post the most inflammatory content need not be socially prominent, and one might even argue that it is the most marginal users who have the strongest incentive to post the most inflammatory content to gain more credit and visibility in their online community~\cite{de2008stormfront}. As a consequence, a lot of energy is spent in removing users who have very little effect on the network structure. The corollary of this result is that communication properties remain almost untouched while cutting off a large amount of extreme emotional content. 

However, combining the results from the analysis of the two strategies, we obtain a viable proposal for an effective strategy of attack, namely combining topological and content information to define the criterion for removal. If removal is carried out by targeting the more inflammatory users above a minimal degree threshold, network dismantling proves to be significantly more effective. A smart combination of topological and content criteria may therefore provide the basis for a new generation of attack strategies that can inform new approaches to the governance of online platforms. 

The role of moderation as a key component of governance remains ambiguous. However, our analysis allow us to provisionally conclude that sentiment based censorship can be a suitable strategy to implement when the objective is not to dismantle the communication network but to maintain the potential exchange of information across the network while limiting the sharing of dysfunctional content. This appears to be possible because extreme users social relevance is very weak in both platforms analyzed.

\section{Methods}\label{sec:Methods}

\subsection{Data}

 We collected messages from both Twitter and Gab, selecting a set of keywords that refer to Donald Trump, including `trump' or `potus'. In the Gab database there are 447,965 messages, spanning a time window of 3 months, from Wednesday 1 August 2018 0:53:10 (GMT) to Monday 29 October 2018 3:03:58 (GMT). In the Twitter database (accessed via the public API) we have 44,934,988 messages, spanning a time window of 2 months, from Sunday 26 August 2018 17:28:20 (GMT) to Saturday 27 October 2018 18:24:37 (GMT).

\subsection{Sentiment analysis}

We used emotional classification of texts to estimate sentiment spectra of single user's corpus of messages. To this aim, we performed text-based sentiment analysis by employing widespread algorithms to classify user’s text corpus. Note that for each message, pre-processing steps were carried out to normalize various aspects of the text before analyzing it (all text have been lower-cased, “broken” unicode fixed).

The text classification employed in the main text is divided in three dimensions, namely valence, arousal and dominance~\cite{warriner2013norms}, where the first is related to the pleasantness, the second to the intensity of emotion and the latter is related to the degree of control exerted by a specific word. To make our analysis more general and flexible, we only include lemmas (that is the base form of the word) in our database. Emotional ratings of about 14000 English lemmas, with scale ranges from 1 (happy, excited , controlled)  to 9 (unhappy, annoyed, dominant), are contained in our database.

To further support our results, we have repeated the analysis --- reported in the Supplemental Material --- with other 3 algorithms, that we detail next. One uses the Big Five personality traits model~\cite{schwartz2013personality, agarwal2014personality}, that describes personality as a vector of five values corresponding to bipolar traits, namely Extroversion, Neuroticism, Agreeableness, Conscientiousness and Openness. Personality detection from text was performed using a database containing ratings of about 1000 lemmas or short sentences. 

Another method aiming at measure the perceived individual well-being was also performed \cite{Schwartz2016, Forgeard2011}. The overall satisfaction present in the text is estimated through multiple components of well-being, that were measured as separate, independent dimensions: Positive Emotions, Engagement, Relationships, Meaning and Accomplishment (PERMA). For each dimension, there is a further specification of its positive and negative acceptation, resulting in 10 different sentiment dimensions.

Finally, the last method we have used to characterize corpus sentiment is based on the eight emotions postulated by Plutchik
\cite{PLUTCHIK1980}, namely Acceptance, Anger, Anticipation, Disgust, Joy, Fear, Sadness and Surprise, that are considered to be the basic and prototypical emotions \cite{Mohammad2013}.

\bibliography{biblio.bib}{}

\begin{thebibliography}{10}
\bibitem{PercolationClarification}
The theory of percolation assumes that we deal with infinite systems, hence the
  percolation phase transition and the percolation point are well-defined only
  in this regime. In a finite system, though, the size of the largest connected
  component is exactly $0$ only when all nodes are removed. Although the drop
  from $S \sim \mathcal{O}(1)$ to $S \simeq 0$ tends to be relatively abrupt
  and localized in finite systems, there is no general criterion to identify
  the percolation point directly from the size of the LCC. An alternative and
  accurate method, which is the one we employ, is to identify the percolation
  point with the fraction of removed nodes for which the size of the second
  largest connected component is maximum.
\bibitem{AssortativityClarification}
Equation~\ref{eq:e1} is not a useful expression when it comes to compute the
  assortativity coefficient directly from degree sequences. We have used,
  instead, a much more convenient expression, namely, $r = \frac{\sum_{ij}
  (A_{ij} - k_i k_j/2m)k_i k_j }{\sum_{ij} (k_i\delta_{ij} - k_i k_j/2m)k_i
  k_j}$.
\end{thebibliography}
\bibliographystyle{plain}

\newpage

\pagebreak
\section*{Supplemental Material}

\renewcommand\thefigure{SM.\arabic{figure}}
\setcounter{figure}{0}
In Figure~\ref{fig:figSM1} we plot the cluster size distribution at percolation point for the 4 networks analyzed in the main text. Figure~\ref{fig:figSM1} is related to the bubble plots of Figures~1B and 1C in the main text.

In Figures~\ref{fig:figSM2}--\ref{fig:figSM4} we display the same analyses as the ones conducted in Figure 4 in the main text. That is, we show the correlation between the degree and the sentiment, the latter being computed using the Big Five algorithm (Fig.~\ref{fig:figSM2}), the Sentiments algorithm (Fig.~\ref{fig:figSM3}) and the Wellbeing algorithm (Fig.~\ref{fig:figSM4}). Each plot is accompanied by the robustness profiles, implemented with sentiment-based attacks. All these curves are qualitatively the same, independent of the algorithm employed to detect the sentiment.

\begin{figure}[t]
    \centering
    \includegraphics[width=0.6\textwidth]{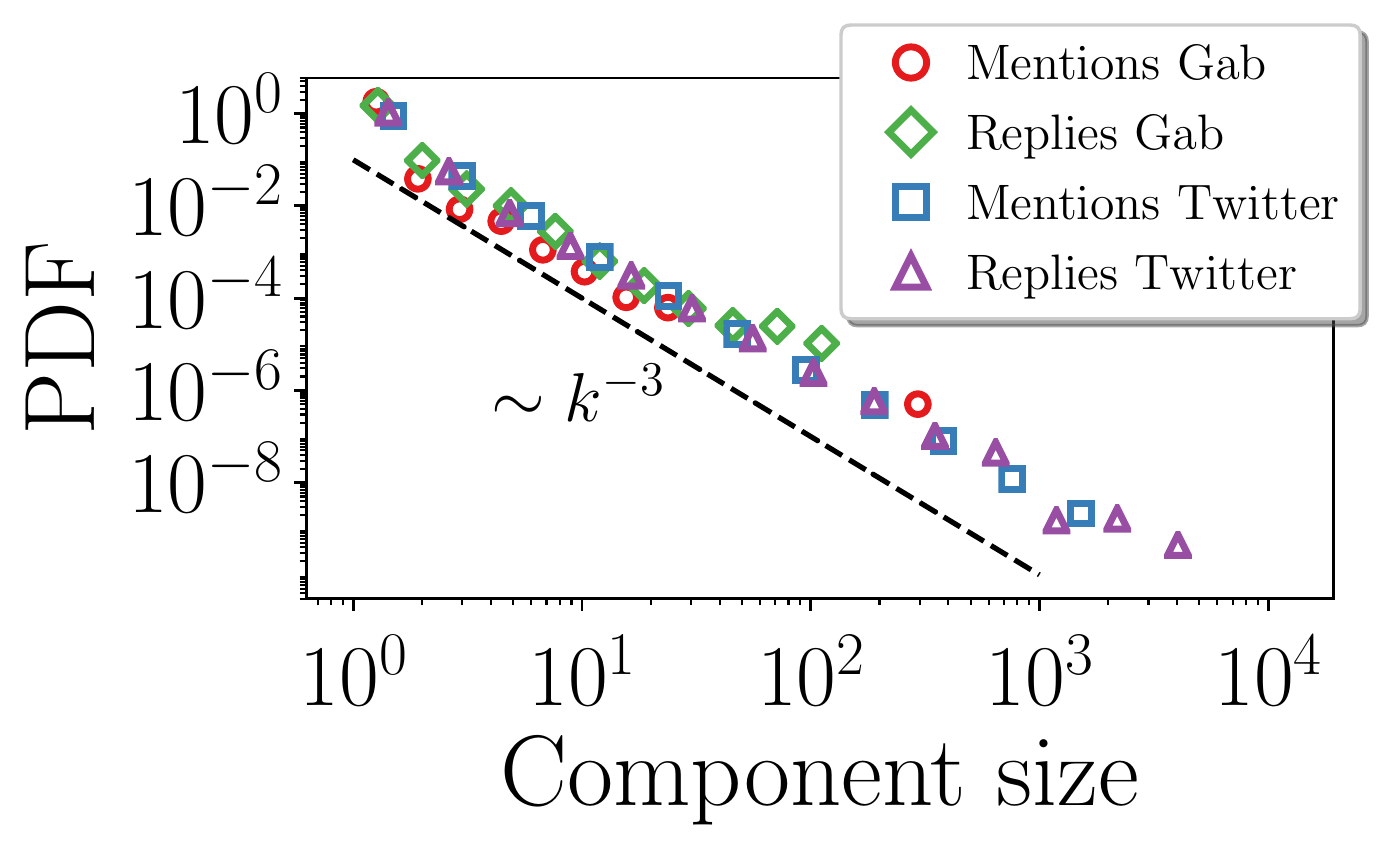}

    \caption{Component size distributions at the percolation points of Figures~1B and 1C. The dashed line is not a fit, but is drawn to show that the distributions decay with an exponent smaller than 3, i.e., the second moment of the distribution diverges.}
    \label{fig:figSM1}
\end{figure}

\begin{figure}[b]
    \centering
    \includegraphics[width=1.\textwidth]{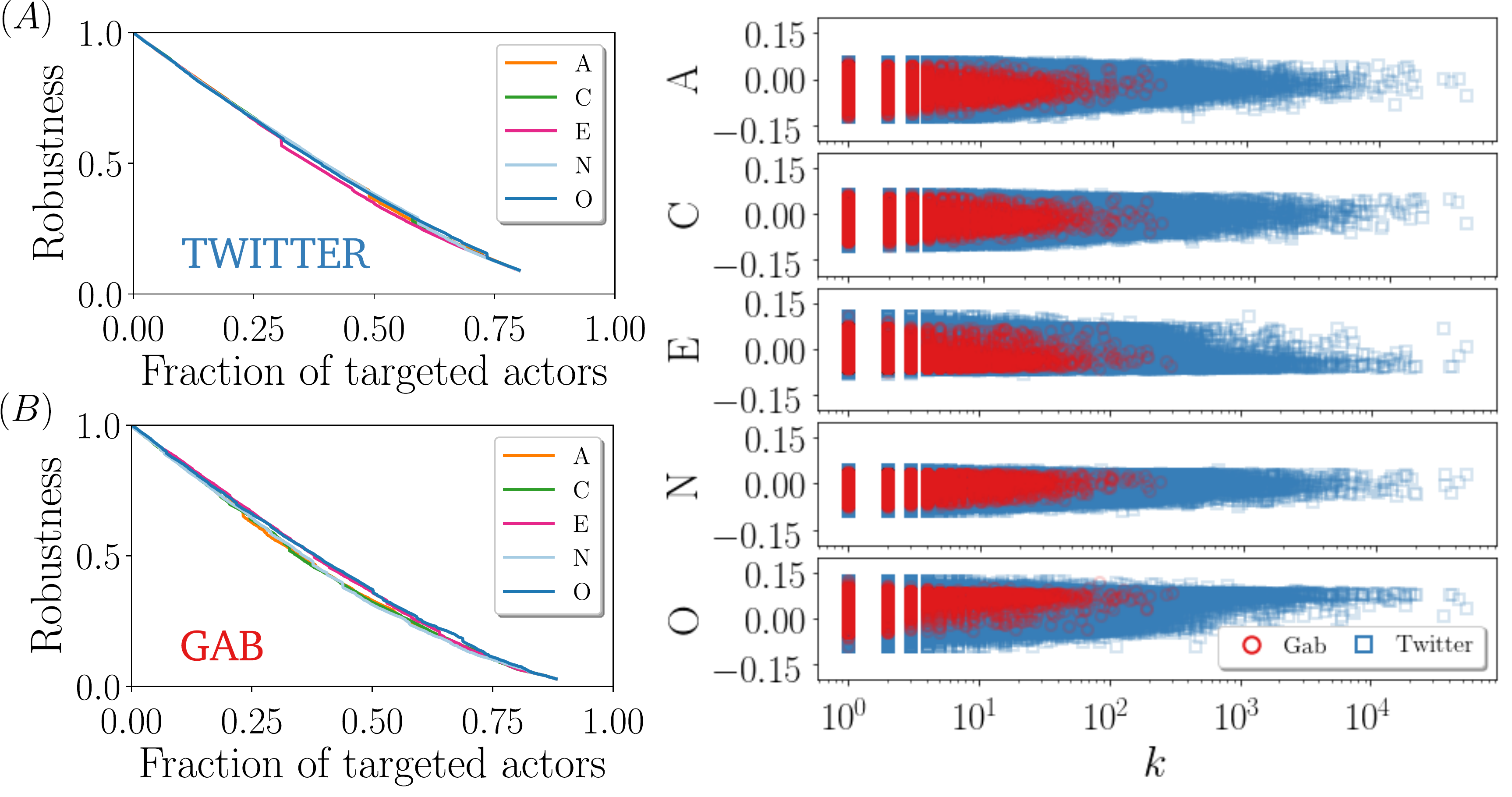}
    \caption{Size of the largest connected component for the sentiment-based attacks in Twitter and in Gab. Each curve corresponds to the sentiment targeted for the removal. On the right, the correlation between the degree of the nodes and the sentiment score computed by using the Big Five algorithm. The corresponding sentiment is written in the vertical axes. For the sake of clarity, in Twitter networks we only plot a random $20\%$ selection of all users.}
    \label{fig:figSM2}
\end{figure}

\begin{figure}
    \centering
    \includegraphics[width=1.\textwidth]{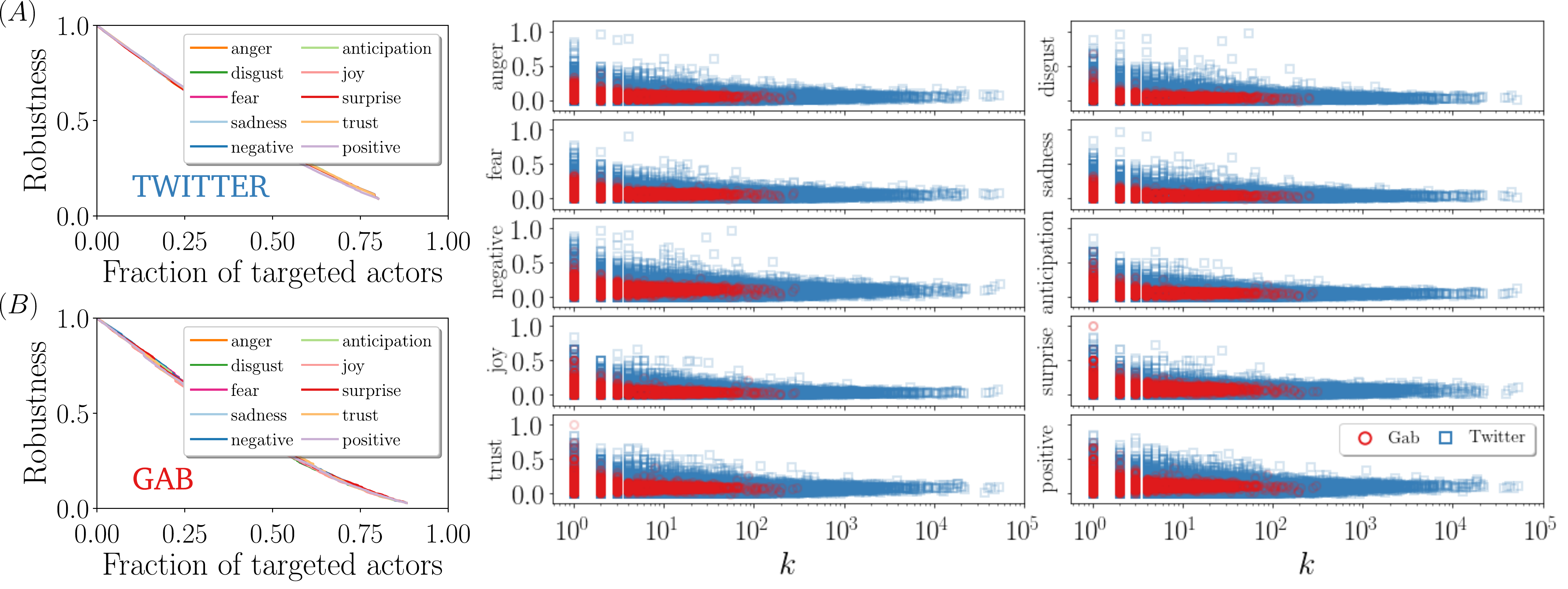}
    \caption{Size of the largest connected component for the sentiment-based attacks in Twitter and in Gab. Each curve corresponds to the sentiment targeted for the removal. On the right, the correlation between the degree of the nodes and the sentiment score computed by using the Sentiments algorithm. The corresponding sentiment is written in the vertical axes. For the sake of clarity, in Twitter networks we only plot a random $20\%$ selection of all users.}
    \label{fig:figSM3}
\end{figure}

\begin{figure}
    \centering
    \includegraphics[width=1.\textwidth]{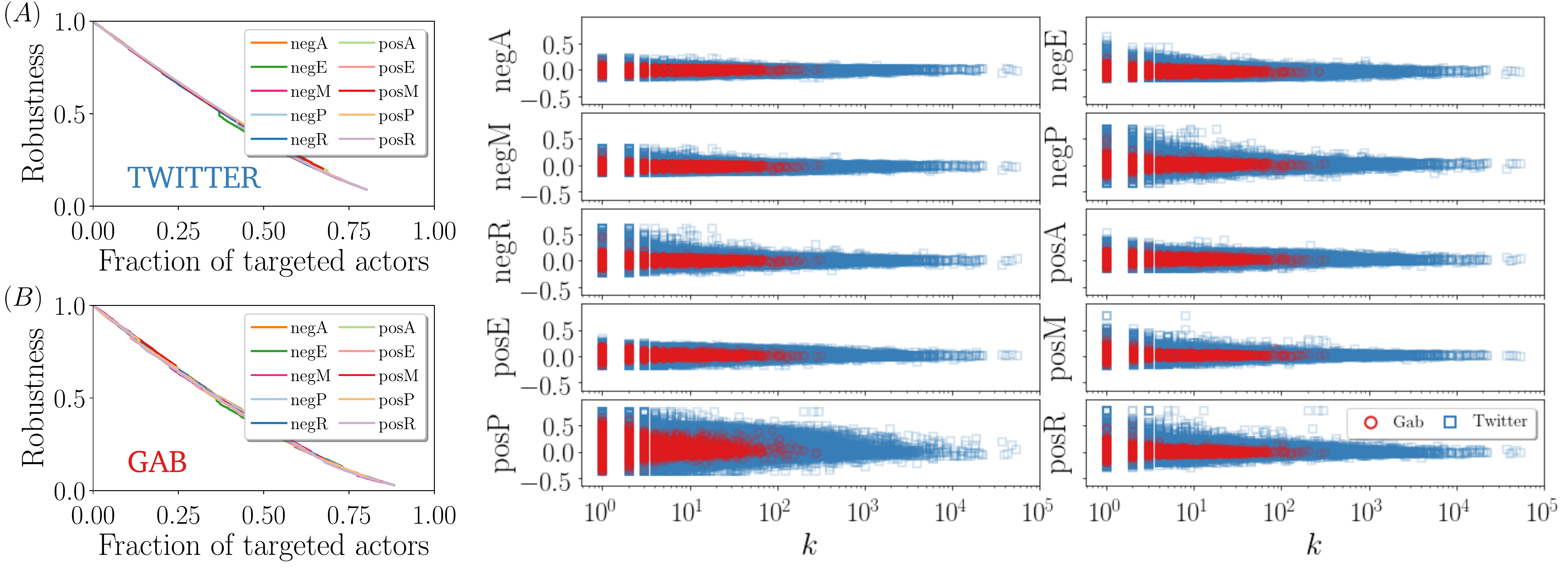}
    \caption{Size of the largest connected component for the sentiment-based attacks in Twitter and in Gab. Each curve corresponds to the sentiment targeted for the removal. On the right, the correlation between the degree of the nodes and the sentiment score computed by using the Wellbeing algorithm. The corresponding sentiment is written in the vertical axes. For the sake of clarity, in Twitter networks we only plot a random $20\%$ selection of all users.}
    \label{fig:figSM4}
\end{figure}

\end{document}